\documentclass{sig-alternate-10pt}
\usepackage[font=bf]{caption}

\newcommand{\subparagraph}{}

\usepackage[compact]{titlesec}

\usepackage{multirow}
\usepackage[sort,nocompress]{cite}
\usepackage[usenames,dvipsnames]{color}
\usepackage{amsfonts}
\usepackage{times}
\usepackage{xspace} \usepackage{epsfig}
\usepackage{amsmath}
\usepackage{amssymb}
\usepackage[hyphens]{url}
\usepackage{listings}
\usepackage{fancyvrb}
\usepackage{graphicx}
\usepackage{subfigure}
\usepackage[english]{babel}
\usepackage{courier}
\usepackage{fnpos}

\usepackage{footnote}
\makesavenoteenv{tabular}
\makesavenoteenv{table}
\usepackage{listings}
\VerbatimFootnotes

\newcommand{\tightcaption}[1]{\vspace{-0.25cm}\caption{#1}\vspace{-0.45cm}}
\newcommand{\tightsection}[1]{\vspace{-0.05cm}\section{#1}\vspace{-0.1cm}}
\newcommand{\tightsubsection}[1]{\vspace{-0.05cm}\subsection{#1}\vspace{-0.1cm}}

\newcommand{\comment}[1]{}
\newcommand{\mycomment}[1]{}
\newcounter{note}[section]

\newcommand{\Section}{\S}

\usepackage{pifont}

\newcommand{\myparatight}[1]{\smallskip\noindent{\bf {#1}:}~}

\newcommand{\name}{{DDA}\xspace}

\newcommand{\session}{\ensuremath{\mathit{s}}}

\newcommand{\QualitySummary}{\ensuremath{\mathit{Median}}}
\DeclareMathOperator*{\argmin}{arg\,min}

\newcommand{\Error}[2]{\ensuremath{\mathit{Err(#1,#2)}}}
\newcommand{\Pred}[1]{\ensuremath{\mathit{Pred(#1)}}}
\newcommand{\Throughput}[1]{\ensuremath{\mathit{#1_{w}}}}

\newcommand{\optimal}{\ensuremath{\mathit{*}}}
\newcommand{\Est}{\ensuremath{\mathit{Est}}}
\newcommand{\Estimation}[1]{\ensuremath{\mathit{\Est(#1)}}}
\newcommand{\Model}{\ensuremath{\mathit{M}}}
\newcommand{\Agg}[2]{\ensuremath{\mathit{Agg(#1,#2)}}}

\newcounter{packednmbr}

\newenvironment{packedenumerate}{\begin{list}{\thepackednmbr.}{\usecounter{packednmbr}\setlength{\itemsep}{0.5pt}\addtolength{\labelwidth}{-4pt}\setlength{\leftmargin}{\labelwidth}\setlength{\listparindent}{\parindent}\setlength{\parsep}{1pt}\setlength{\topsep}{0pt}}}{\end{list}}

\newenvironment{packeditemize}{\begin{list}{$\bullet$}{\setlength{\itemsep}{0.5pt}\addtolength{\labelwidth}{-4pt}\setlength{\leftmargin}{\labelwidth}\setlength{\listparindent}{\parindent}\setlength{\parsep}{1pt}\setlength{\topsep}{0pt}}}{\end{list}}

\begin{document}

%\title{Cross-Session Throughput Prediction for Initial Video Bitrate Selection}
\title{\name: Cross-Session Throughput Prediction with Applications to Video Bitrate Selection}

%\author{TBD}
\author{Junchen Jiang, Vyas Sekar\\CMU, USA
\and
Yi Sun\\ICT, China}

\maketitle

\begin{abstract}

User experience of video streaming could be greatly improved by selecting a high-yet-sustainable initial video bitrate, and it is therefore critical to accurately predict throughput before a video session starts.
%Accurate prediction of throughput is crucial for video streaming to optimize user experience.
% by picking a highest bitrate below the predicted throughput. Ideally, we would like to predict the throughput of a video session when it starts. However, existing approaches either require substantial historical throughput measurement collected from the same client, which is often unavailable, or waist the first few video chunks (e.g., tens of seconds) to warm-up before the player can predict the throughput. 
Inspired by previous studies that show similarity among throughput of similar sessions (e.g., those sharing same bottleneck link), we argue for a {\em cross-session} prediction approach, where throughput measured on sessions of different servers and clients is used to predict the throughput of a new session. 
%This approach has recently become plausible as many video service providers begin to constantly collect throughput measurements from video players. 
%Despite of throughput similarity among sessions of spatial and temporal correlation, it is, however, much more challenging to turn such similarity into accurate prediction than it first appears. 
In this paper, we study the challenges of cross-session throughput prediction, develop an accurate throughput predictor called \name, and evaluate the performance of the predictor with real-world datasets. We show that \name predicts throughput more accurately than simple predictors and conventional machine learning algorithms; e.g., \name's 80\%ile prediction error of \name is $\geq$ 50\% lower than other algorithms. We also show that this 
improved accuracy enables video players to select a higher sustainable initial bitrate; e.g., compared to initial bitrate without prediction, \name leads to $4\times$ higher average bitrate.

\end{abstract}

\tightsection{Introduction}
\label{sec:intro}

Many Internet applications can benefit from estimating the client-server
throughput. For instance, accurate estimation of throughput helps content
distribution networks to redirect clients to servers that provide the best
performance. Similarly, peer-to-peer networks select the best peers based on
the estimation of their throughput performance.

Our focus in this paper is on the initial video bitrate selection when a video
player starts. A video player should ideally pick the highest initial bitrate
that is sustainable (i.e., below the throughput), in order to ensure desired user experience of video
streaming. Existing approaches to initial bitrate selection, however, are
inefficient. Table~\ref{tab:video} shows measured anecdotal evidence of such
inefficiencies from several commercial providers.  Fixed-bitrate players that
use the same bitrate for the whole video session often intentionally use low
bitrate to prevent mid-stream rebuffering (e.g., NFL, Lynda). Even if bitrate
can be adapted midstream (e.g.,~\cite{dash,netflix,festive}) the player often
conservatively starts with a low bitrate and takes a significant time to reach
the optimal bitrate (e.g., Netflix).  Furthermore, for short video clips such
adaptation may not reach the desired bitrate before the video finishes (e.g., Vevo music clips).

\begin{table}[t!]
\begin{footnotesize}
    \begin{tabular}{p{1.2cm}|p{1.5cm}|p{2.1cm}|p{2.4cm}}
    {\bf Streaming protocol}   & {\bf Examples} & {\bf Limitations} & {\bf How throughput prediction helps} \\ \hline\hline
    Fixed bitrate & NFL,Lynda, NYTimes        & Too low bitrate, a few chunks are & Higher bitrate with no re-buffering or \\ \cline{1-2}
    Adaptive bitrate & ESPN,Vevo, Netflix      &  wasted to probe throughput & long startup time
    \end{tabular}
\end{footnotesize}
\tightcaption{Limitations of today's video players and how they benefit from throughput prediction.
www.lynda.com  uses fixed bitrate of 520Kbps (360p) by default.  Netflix (www.netflix.com/WiMovie/70136810?tr kid=439131) takes roughly 25 seconds to adapt from the initial bitrate (560Kbps) to the highest sustainable bitrate (3Mbps).}
\label{tab:video}
\end{table}

The importance of initial bitrate selection (e.g., avoid users quitting)
naturally makes a case for a predictive approach -- predicting the TCP
throughput before a session starts.  Inspired by prior work on shared
measurements~\cite{seshan1997spand}, we explore a {\em cross-session} approach where the
TCP throughput of other sessions is used to predict the throughput of a new
session.  Intuitively, we want to build a prediction model for each
client-server pair as a function of key session features available to us (e.g.,
ISP, connection type).  This has the natural advantage that it incurs no
additional measurement overhead and leverages all available sessions even if
there is no history between same client and server.  While this idea is not
particularly new, we believe that revisiting this is timely in light of the
need for video quality optimization and the availability of large-scale
throughput measurements to many video service providers.\footnote{There has
been surprisingly little work in exploring this idea for throughput prediction
since the early work of \cite{seshan1997spand}.}

However, it is challenging to predict throughput accurately based on other
sessions' throughput, because of a complex underlying interaction between
session features and throughput. There are two manifestations of this
complexity (see \Section\ref{sec:analysis} for more details).  First,
throughput usually can only be accurately predicted by combination of multiple
features.  For instance, only sessions from a particular ISP-server-device
combination may have similarly low throughput, but the individual ISP, server
or device may manifest no problem.  Second,  the best feature combination to
predict throughput  differs across sessions.  For instance, for sessions in one
ISP, the best feature to predict their throughput is last-mile connection
(e.g., last hope is the bottleneck), while for those in another ISP, the best
feature is time of day (e.g., due to a strong diurnal pattern).

To address these challenges, we present \name (Data-Driven Aggregation), which predicts each session's throughput by
 an expressive prediction model that captures temporal similarity (e.g., sessions happening
within a smaller time window) and spatial similarity (e.g., sessions matching
more features with the session under prediction) between previously observed sessions and
the session under prediction. Such prediction models allow \name to predict throughput accurately by aggregating sessions with similar throughput. Instead of using a single prediction
model, \name customizes the prediction model for each session under prediction.
To pick the prediction model that yields high prediction accuracy for each
session, \name adopts a data-driven approach and learns the best prediction
model by searching for the best prediction model to similar existing history
sessions.

%To tackle the challenges, one may be tempted to use sessions matching all features and happening in the same time with the session under prediction. However, it is seldom to have enough session for prediction. Therefore, we have to address a {\em space-time tradeoff} between higher spatial correlation (e.g., using sessions of the same client and server in a longer history) and higher temporal correlation (e.g., using sessions between different clients or servers but in a shorter history).

%On a high level, \name strikes this space-time tradeoff by leveraging the insight that the best space-time tradeoff of a session is persistent. This suggests that \name can learn the best space-time tradeoff for each session from similar history sessions. Once the best space-time tradeoff is learned, \name uses it to find sessions with similar throughput to the session under prediction. \jc{add more intuition of \name}

In summary, this paper makes three key contributions. 
\begin{packedenumerate}
\item First, we use a dataset of real-world throughput measurement of 9.9 million sessions to show the challenges of cross-session throughput prediction (\Section\ref{sec:analysis}). 
\item Second, we present a concrete cross-session throughput predictor, \name that addresses the above challenges (\Section\ref{sec:algorithm}). 
\item Finally, our evaluation based on two real-world datasets shows that \name can predict throughput more accurately than simple predictors and conventional machine learning algorithms, and that due to more accurate throughput prediction, \name allows a video player to select a higher-yet-sustainable initial bitrate (\Section\ref{sec:eval}).

%we demonstrate the benefits of using \name to improve video bitrate selection (\Section\ref{sec:eval}). We show that \fillme.
\end{packedenumerate}
%In summary, our evaluation suggests that \name can improve prediction accuracy over traditional machine learning algorithms by \fillme, and that bitrate selected based on \name is \fillme\% higher than baseline solutions.

\tightsection{Related Work}
\label{sec:related}

At a high-level, our work is related to prior work in measuring Internet path
properties,   bandwidth measurements, and video-specific bitrate selection.
With respect to prior measurement work, our key contribution is showing a
practical data-driven approach for throughput prediction.  In terms of  video,
our predictive approach offers a more systematic bitrate selection mechanism.

\myparatight{Measuring path properties} Studies on path properties have shown
prevalence and persistence of network bottlenecks
(e.g.,~\cite{hu2005measurement}), constancy of various network
metrics~\cite{zhang2001constancy}, longitudinal patterns of cellular
performance (e.g.,~\cite{nikravesh2014mobile}), and spatial similarity of
network performance (e.g.,~\cite{balakrishnan1997analyzing}). While \name is
inspired by these insights, it addresses a key gap because these efforts fall
short of providing a prescriptive algorithm for throughout prediction.

% the shown substantial throughput similarity, but as shown in
%\Section\ref{sec:analysis}, achieving accurate prediction is hard because of
%complex underlying interaction between sessions' throughput.

%%Existing approaches to
%prediction of bandwidth-related metrics (e.g., available bandwidth, throughput)
%%generally fall into three categories. The most similar approach to ours is
%shared prediction.

\myparatight{Bandwidth measurement}  Unlike prior ``path mapping'' efforts
(e.g.,~\cite{madhyastha2006iplane, seshan1997spand,ramasubramanian2009treeness,
dabek2004vivaldi}), \name uses a data-driven model based on available session
features (e.g., ISP, device). Specifically,  video  measurements are taken
within a constraint sandbox environment (e.g., browser) that do not offer interface for path
information (e.g., traceroute).  Other approaches use  packet-level probing to
estimate the end-to-end performance metrics (e.g.,~\cite{prasad2003bandwidth,
hu2004locating, strauss2003measurement, jain2003end}).  Unlike \name, these
require additional measurement and often need full client/server-side control which
is often infeasible in the wild.  A third class of approaches leverages the
history of the  same client-server pair (e.g.,~\cite{vazhkudai2001predicting,
jain2005end, swany2002multivariate, mirza2007machine, he2005predictability}).
However, they are less reliable when the available history of the same client
and server is sparse. 

%Third approach, which is the most similar to ours, is to predict the
%performance between a host and server using the information between other
%hosts and servers. However, they either rely on network topology to correlate
%measurements (e.g., ~\cite{madhyastha2006iplane, seshan1997spand}), or fall
%short of predicting throughput (e.g.,~\cite{ramasubramanian2009treeness,
%dabek2004vivaldi}), while \name relies on only session features to
%automatically discover sesssion with similar throughput and achieve accurate
%prediction. 

%Similar to
%\name, predictive bitrate selection picks bitrate based on prediction of
%throughput. However,

\myparatight{Bitrate selection} Choosing high and sustainable bitrate is
critical to video quality of experience~\cite{sigcomm13athula}.  
 Existing methods (e.g.,~\cite{miller2015low,festive})
require either history measurement between the same client and server or the
player to probe the server to predict the throughput. In contrast, \name is able to predict
throughput before a session starts. 
%In this perspective, video control plane
%(e.g.,~\cite{sigcomm12conviva,c3}) is similar to \name, but it fails to provide
%details on how they correlate sessions with similar quality and falls short of
%predicting throughput directly.  
 Other approaches include switching bitrate midstream (e.g.,~\cite{huang2014buffer,tian2012towards,yin2014toward})
 but do not focus on the initial bitrate problem which is the focus of \name.

\tightsection{Datasets}
\label{sec:dataset}

We use two datasets of HTTP throughput measurement to evaluate \name's performance: (i) a primary dataset collected by FCC's Measuring Broadband American Platform~\cite{fcc-2014} in September 2013, and (ii) a supplementary dataset collected by a major VoD provider in China. 

\begin{figure}[t!]
\centering
%\vspace{-0.4cm}
\includegraphics[width=0.4\textwidth]{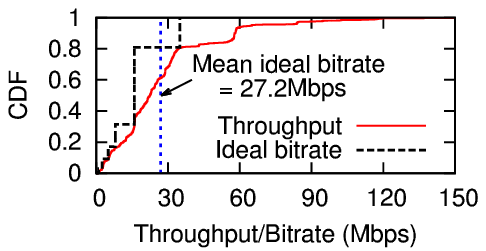}
\tightcaption{Distribution of throughput in the FCC dataset}
\label{fig:cdf-throughput}
\end{figure}

\myparatight{FCC dataset} This dataset consists of 9.9 million sessions and is collected from 6204 clients in US spanning 17 ISPs. In each test, a client set up an HTTP connection with one of the web servers for a fixed duration of 30 seconds and attempted to download as much of the payload as possible. It also recorded average throughput at 5 second intervals during the test. 
The test used three concurrent TCP connections to ensure the line was saturated. 
%Each connection used in the test counted the numbers of bytes transferred and was sampled periodically by a controlling thread. The sum of these counters divided by the time elapsed was taken as the total throughput of the line.
%\footnote{Factors such as TCP slow start and congestion were taken into account by repeatedly transferring small chunks (256 kB) of the target payload before the real testing began. This ``warm up'' period was said to have been completed when three consecutive chunks were transferred at within 10\% of the speed of one another. All three connections were required to have completed the warm up period before the timed testing began. The ``warm-up'' period was excluded from the measurement results.}
Reader may refer to~\cite{fcc-methodology} for more details on the methodology. 

Figure~\ref{fig:cdf-throughput} shows the throughput distribution of all sessions. It also shows the distribution of ideal bitrate (i.e., highest bitrate chosen from \{0.016, 0.4, 1.0, 2.5, 5.0, 8.0, 16.0, 35.0\}Mbps\footnote{The bitrates are recommended upload encoding by YouTube~\cite{youtube-bitrates,wiki-bitrates}.} below the throughput). With perfect throughput prediction, we should be able to achieve average bitrate of 26.9Mbps with no session suffering from re-buffering. Compared to the fixed initial bitrate (e.g., 2.5Mbps) used today, this suggests a large room of improvement.

The clients
%\footnote{In the FCC dataset, each client has a fixed ISP, State and Technology.} 
represent a wide spatial coverage of ISPs, geo-locations, and connection technology (see Table~\ref{tab:fcc-stats}). Although the number of targets are relatively small, the setting is very close to what real-world application providers face -- the clients are widely distributed while the servers are relatively fewer. 
%While the spatial coverage of FCC dataset naturally imitates a real-world dataset collected by a video provider, 
In addition, its measurement frequency (i.e., each client fetching content from each server once every hour) provides a unique opportunity to test the prediction algorithms' sensitivity to different measurement frequency. 
For instance, to emulate the effect of reduced data, we take one (the first) 5-second throughput sample from each test, and then randomly drop (e.g., 90\% of) the available measurements to simulate a dataset where each client accesses a server less frequently (e.g., in average once every 10 hours).

%To make it realistic, we first take one (the first) 5-second throughput sample from each test, and then we randomly drop (say, 90\%) of the history measurement of each client to simulate a dataset where each client access a server, say, in average every 10 hours. Such random dropping allows us to evaluate the prediction algorithms' sensitivity against measurement frequency, where a more robust prediction algorithm should be impacted less significantly by a higher drop rate than other algorithms.

%Besides client-side coverage, the two applications (web downloading and video streaming) represent two different types of traffic: small burst traffic (in average \fillme bytes) and long traffic (in average \fillme bytes). In \Section\ref{sec:eval}, we will discuss the different performance of various algorithms on the two applications.

\begin{table}[t]
\begin{footnotesize}
    \begin{tabular}{p{1.4cm}|p{5.0cm}|p{1.5cm}}
 {\bf Feature} & {\bf Description}                        & {\bf \# of unique values} \\ \hline\hline
    ClientID   & Unique ID associated to a client         & 6204                \\
    ISP        & ISP of client (e.g., AT\&T)              & 17                  \\
    State      & The US state where the client is located & 52                  \\
    Technology & The connection technology (e.g., DSL)   & 5                   \\
    Target     & The server-side identification     & 30 \\
    Downlink   & Advertised download speed of the last connection (e.g., 15MB/s) & 36 \\
    Uplink     & Advertised upload speed of the last connection  (e.g., 5MB/s)   & 25 \\
    \end{tabular}
\end{footnotesize}
\tightcaption{Basic statistics of the FCC dataset.}
\vspace{-0.2cm}
\label{tab:fcc-stats}
\end{table}

\myparatight{Supplementary VoD dataset} As a supplementary dataset, we use throughput dataset of 0.8 millions VoD sessions, collected by a major video content provider in China. Each video session has the average throughput and a set of features, that are different from the FCC dataset, including content name, user geolocation, user ID and server IP. This provides an opportunity to test the sensitivity of the algorithms to different sets of available features. %However, the video dataset lacks some key features, which results in lower prediction accuracy (\Section\ref{sec:eval}); for instance, the dataset only has two days and no last-connection information, so we have no access to longitudinal features (e.g., time-of-day) or connectivity features (e.g., WiFi or cable)\footnote{Note that such features are available in general (e.g., in~\cite{imc-akamai,conext13shedding})}. 
\tightsection{Simple Predictors are Not Sufficient}
\label{sec:analysis}

This section starts by showing that simple predictors fail to yield desirable prediction accuracy, and then shows fundamental challenges of cross-session throughput prediction. 

\begin{packeditemize}
\item First, we consider the {\bf last-mile predictor}, which uses sessions with the same $\mathit{downlink}$ feature (see definition in Table~\ref{tab:fcc-stats}) to predict a new session's throughput.  This is consistent to the conventional belief that last-mile connection is usually the bottleneck. However, Figure~\ref{fig:last-mile-1} and \ref{fig:last-mile-2} show substantial prediction error\footnote{Given throughput prediction $p$ and actual throughput $q$, we define four types of prediction error: non-normalized absolute prediction error: $|p-q|$, normalized absolute prediction error: $\frac{|p-q|}{q}$, non-normalized signed prediction error: $p-q$, normalized signed prediction error: $\frac{p-q}{q}$.}, especially on the tail where at least 20\% of sessions have more than 20\% error (Figure~\ref{fig:last-mile-2}). To put it into perspective, if a player chooses bitrate based on throughput prediction that is 20\% higher or lower than the actual, the video session will experience mid-stream re-buffering or under-utilize the connection. Finally, Figure~\ref{fig:last-mile-3} and \ref{fig:last-mile-4} show that the prediction error is two-sided, suggesting that simply adding or multiplying the prediction by a constant factor will not fix the high prediction error.

\begin{figure}[h!]
\centering
\vspace{-0.5cm}
\subfigure[Non-normalized absolute]
{
        \includegraphics[width=0.22\textwidth]{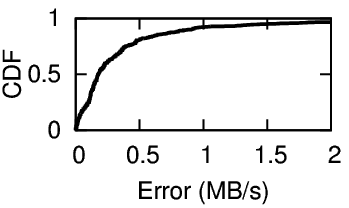}
        \label{fig:last-mile-1}
}
\subfigure[Normalized absolute]
{
        \includegraphics[width=0.22\textwidth]{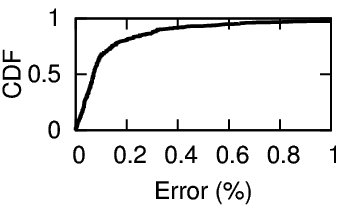}
        \label{fig:last-mile-2}
}\\
\vspace{-0.5cm}
\subfigure[Non-normalized signed]
{
        \includegraphics[width=0.22\textwidth]{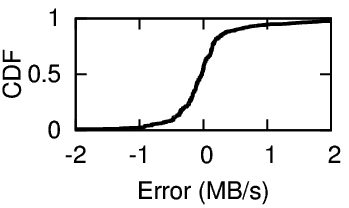}
        \label{fig:last-mile-3}
}
\subfigure[Normalized signed]
{
        \includegraphics[width=0.22\textwidth]{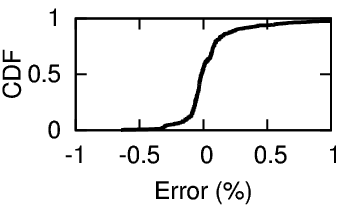}
        \label{fig:last-mile-4}
}
\tightcaption{Prediction error of the last-mile predictor}
\label{fig:last-mile}
\end{figure}

\item Second, we consider the {\bf last-sample predictor}, which uses the throughput of the last session of the same client-target pair to predict the throughput of a future session. 
However, the last-sample predictor is not reliable as the last sample is too sparse and noisy to offer reliable and accurate prediction. 
Figure~\ref{fig:last-sample} shows that, similar to the last-mile predictor, (i) the prediction error, especially on the tail, is not desirable -- more than 25\% of sessions have more than 20\% normalized prediction error, and (ii) the prediction error is two-sided, suggesting the prediction error cannot be fixed by simply adding or multiplying the prediction with a constant factor.

\begin{figure}[h!]
\centering
\vspace{-0.5cm}
\subfigure[Non-normalized absolute]
{
        \includegraphics[width=0.22\textwidth]{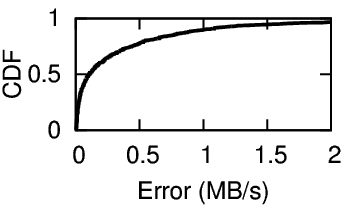}
        \label{fig:last-sample-1}
}
\subfigure[Normalized absolute]
{
        \includegraphics[width=0.22\textwidth]{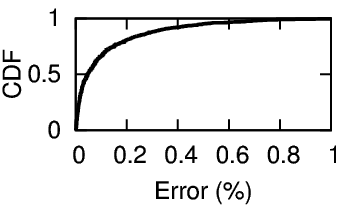}
        \label{fig:last-sample-2}
}\\
\vspace{-0.5cm}
\subfigure[Non-normalized signed]
{
        \includegraphics[width=0.22\textwidth]{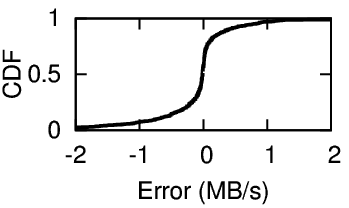}
        \label{fig:last-sample-3}
}
\subfigure[Normalized signed]
{
        \includegraphics[width=0.22\textwidth]{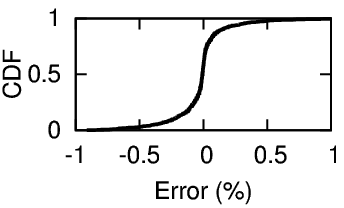}
        \label{fig:last-sample-4}
}
\tightcaption{Prediction error of last-sample predictor}
\label{fig:last-sample}
\end{figure}

\end{packeditemize}

\begin{figure}[t!]
\centering
\subfigure[The average throughput of sessions matching all and a subset of three features: $\mathit{ISP=}$~Frontier, $\mathit{Technology=}$~DSL and $\mathit{Target=}$~samknows1.lax9.level3.net (X). Time: 18:00-00:00 UTC, Oct 7, 2013]
%Example of the effect of feature combinations: only sessions that are in Frontier (X), use DSL (Y) and access samknows1.lax9.level3.net (Z) have low throughput, while those sharing fewer features manifest no problem. Time: 18:00-00:00 UTC, Sep 7, 2013]
{
        \includegraphics[width=0.4\textwidth]{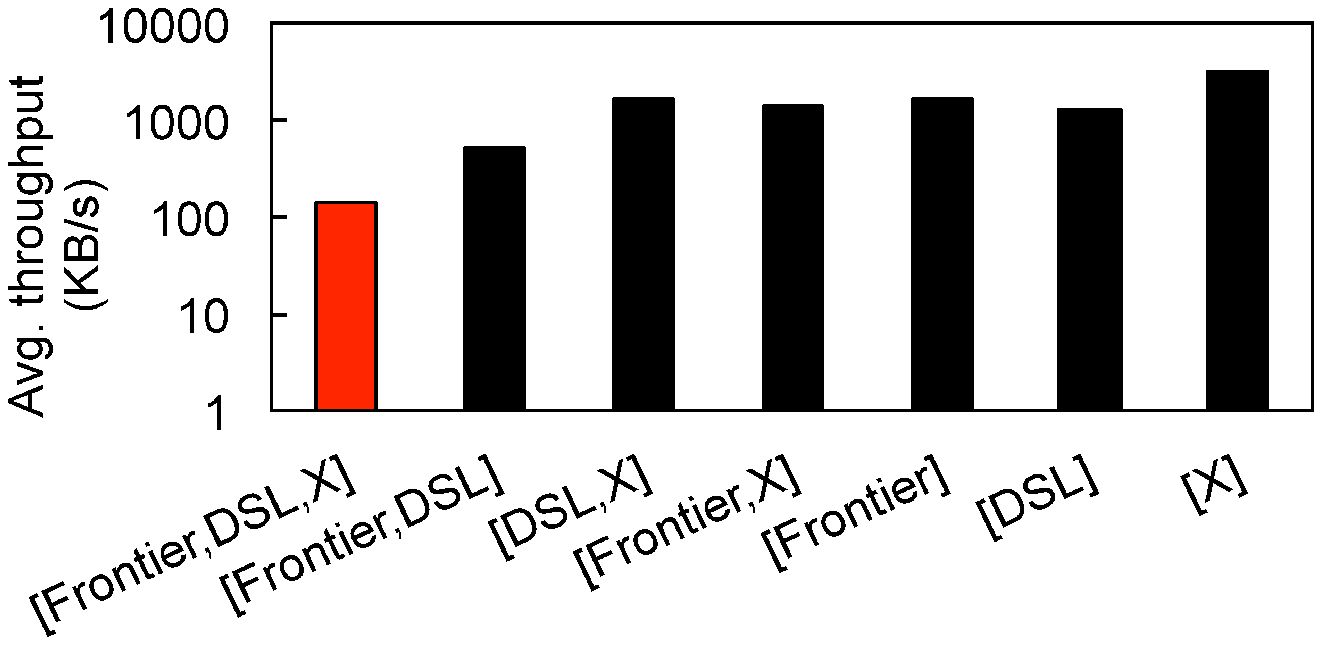}
        \label{fig:high-dimensionality}
}
\\ \vspace{-0.5cm}
\subfigure[The relative information gain of $\mathit{Target}$ in two ISPs over time.]
%Example of same feature having different impact of different sessions: the relative information gain of the same feature (i.e., Technology) varies both across sessions in different ISPs and over time.]
{
        \includegraphics[width=0.4\textwidth]{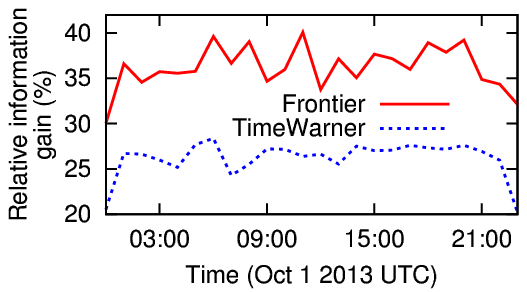}
        \label{fig:diversity}
}
\tightcaption{Two manifestations of the high complex interaction between session features and the throughput.}
\label{fig:challenges}
\end{figure}

%\tightsubsection{Challenges}
%\label{subsec:challenge}

\myparatight{Challenges} The fundamental challenge to produce accurate prediction is the complex underlying interactions between session features and their throughput.
%, which make it challenging to figure out which sessions have similar throughput to the session under prediction.
In particular, there are two manifestations of such high complexity.

%The low prediction accuracy of the two simple solutions suggests a gap between what previous studies show (e.g., for each session, there are {\em some} other sessions with similar throughput) and what we need (i.e., how to identify {\em which} sessions have the similar throughput). Such gap is a result of the complex underlying interactions between session features and their throughput, which make it challenging to figure out which sessions have similar throughput to the session under prediction. In particular, there are manifestations of such high complexity.

%\mypara{High dimensionality} 
First, the simple predictors are both based on single feature (e.g., downlink or time), while combinations of multiple features often have a much greater impact on throughput than individual features. This can be intuitively explained as the throughput is often {\em simultaneously} affected by multiple factors (e.g., the last-mile connection, server load, backbone network congestion, etc), and that means sessions sharing individual features may not have similar throughput. Figure~\ref{fig:high-dimensionality} gives an example of the effect of feature combinations. It shows the average throughput of sessions of ISP Frontier using DSL fetching target samknows1.lax9.level3.net, and average throughput of sessions having same values on one or two of the three features. The average throughput when all three features are specified is at least 50\% lower than any of other cases. Thus, to capture such effect, the prediction algorithm must be expressive to combine multiple features. %Thus, algorithms such as Naive Bayes, as shown in \Section\ref{sec:eval}, could lead to low prediction accuracy.

%\mypara{High diversity} 
Second, the simple predictors both use same feature to all sessions, but the impact of same features on different sessions could be different. For instance, throughput is more sensitive to last-mile connection when it is unstable (e.g., Satellite), and it depends more to ISP during peak hours when the network tends to be the bottlenecks. Figure~\ref{fig:diversity} shows a real-world example. Relative information gain\footnote{$RIG(Y|X) = 1-H(Y|X)/H(Y)$, where $H(Y)$ and $H(Y|X)$ are the entropy of $Y$ and the average conditional
entropy of $Y$~\cite{informationgain}.} is often used to quantify how useful a feature is used for prediction. The figure shows the relative information gain of feature $\mathit{Target}$ on the throughput of sessions in two ISPs over time.
It shows that the impact of the same feature varies across sessions in different hours and in different ISPs. %It is shown that the same feature could have different impact on different sessions and over time.
%Thus, if an algorithm treats each feature (e.g., PCA which assigns a same weight to it) independently to the session under prediction, it could lead to low prediction accuracy.

We will see in \Section\ref{sec:eval} that due to the complex underlying interactions between features and throughput, it is non-trivial for conventional machine learning algorithms (e.g., decision tree, naive bayes) to yield high accuracy.

\tightsection{Predicting Throughput Using \name}
\label{sec:algorithm}

In this section, we present the \name approach that yields accurate throughput prediction (\Section\ref{sec:eval}). 
We start with an intuitive description of \name before formally describing the algorithm.

\tightsubsection{Insight of \name}

At a high level, \name finds for any session $\session$ a {\em prediction model} -- a pair of features and time range, which is used to aggregate history sessions that match the specific features with $\session$ and happened in the specific range. 
%Intuitively, we want the aggregated history sessions to have similar throughput with $\session$ so that they can produce a highly accurate prediction.

\begin{figure}[t!]
\centering
%\vspace{-0.4cm}
\includegraphics[width=0.4\textwidth]{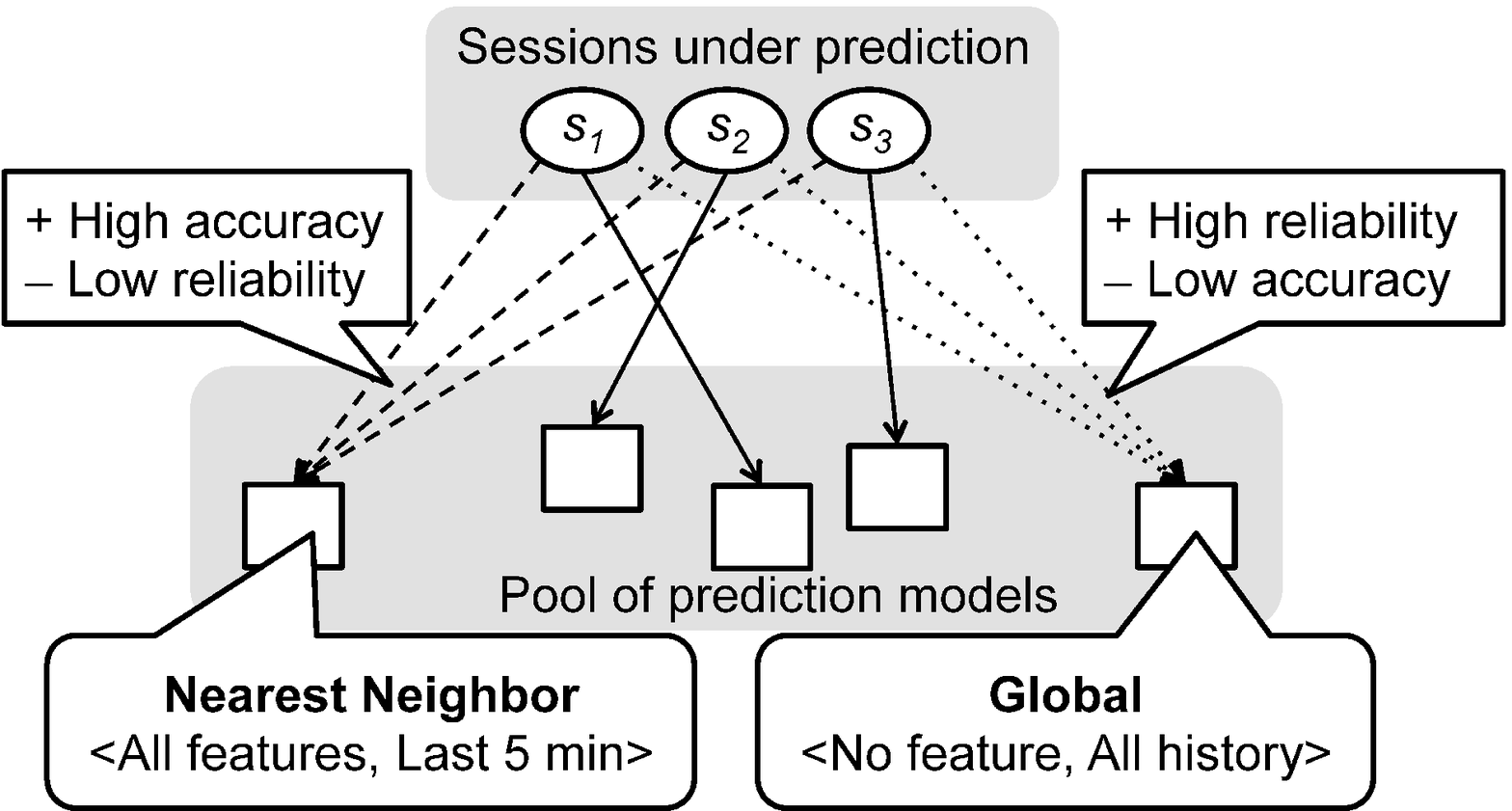}
\tightcaption{Mapping between sessions under prediction and prediction models.}
\label{fig:tbd-motivation}
\end{figure}

To motivate how \name maps a session to a prediction model, let us consider two strawmen of session-model mapping shown in Figure~\ref{fig:tbd-motivation}. 
The first strawman maps each session $\session$ to the ``Nearest Neighbor'' prediction model (dash arrows), which aggregates only history sessions matching all features with $\session$ and happening in very short time (e.g., 5 minute) before $\session$. Theoretically, ``Nearest Neighbor'' model should be highly accurate as it represents sessions that are the most similar to $\session$, but history sessions meeting this requirement are too sparse to provide reliable prediction. 
Alternatively, one can map any $\session$ to the ``Global'' prediction model (dot arrows), which aggregates all history sessions regardless of their features or happening time. While ``Global'' model is highly reliable as it has substantial samples in history, the accuracy is low because it does not capture the effect of feature combination introduced in the last section.

Ideally, we would like achieve both high accuracy and high reliability. To this end, \name (shown by solid arrows in Figure~\ref{fig:tbd-motivation}) differs from the above strawmen in two important aspects. First, \name finds for a given session a prediction model between the Nearest Neighbor and Global prediction models, so that it  strikes a balance between being closer to Nearest Neighbor for accuracy and being closer to Global for reliability. The resulting prediction model should be expressive (e.g., have more features) and yet have enough samples to offer a reliable prediction.
Second, instead of mapping all sessions to the same prediction model, \name maps different sessions to different prediction models, which allows \name to address inherent heterogeneity that the same feature has different impact on different sessions.

\tightsubsection{Design of \name}

\myparatight{Overall workflow} \name uses two steps to predict the throughput of a new session $\session$.
\begin{packedenumerate}
\item First, \name learns a prediction model $\Model^{\optimal}_\session$ based on history data. A prediction model is a pair of feature combination and time range. 
\item Second, \name estimates $\session$'s throughput by the median throughput of sessions in $\Agg{\Model^{\optimal}_\session}{\session}$ that match $\session$ on the feaures of $\Model^{\optimal}_\session$ and are in the time range of $\Model^{\optimal}_\session$. I.e., \name's prediction is $\Pred{\session}=\QualitySummary(\Agg{\Model^{\optimal}_\session}{\session})$.
\end{packedenumerate}

\myparatight{Learning of prediction model} First, \name learns a prediction model $\Model^{\optimal}_\session$ based on history data from a pool of all possible prediction models, i.e., pairs of all feature combinations (i.e., $2^n$ subsets of $n$ features in Table~\ref{tab:fcc-stats}) and possible time windows. Specifically, the possible time windows include time windows of certain history length (i.e., last 10 minutes to last 10 hours) and those of same time of day/week (i.e., same hour of day in the last 1-7 days or same hour of week in the last 1-3 weeks). 

The objective of $\Model^{\optimal}_\session$ is to minimize the prediction error, $\Error{\Pred{\session}}{\Throughput{\session}}=\frac{|\Pred{\session}-\Throughput{\session}|}{\Throughput{\session}}$, where  $\Throughput{\session}$ is the actual throughput of $\session$. That is,

%We start with \name's workflow to predict throughput of session $\session$:
%\begin{packedenumerate}
%\item First, \name picks a prediction model $\Model^{\optimal}_\session$ from a pool of all prediction models, i.e., pairs of all feature combinations (i.e., $2^n$ subsets of $n$ features in Table~\ref{tab:fcc-stats}) and time windows. Specifically, the possible time windows include time windows of certain history length (i.e., last 10 minutes to last 10 hours) and those of same time of day/week (i.e., same hour of day in the last 1-7 days or same hour of week in the last 1-3 weeks). 
%\item Second, once prediction model $\Model^{\optimal}_\session$ is picked for $\session$, \name aggregates history sessions based on $\Model^{\optimal}_\session$. For instance, given $\Model^{\optimal}_\session=\Pair{ISP}{1hr}$, \name will aggregate all history sessions who are in the same ISP with $\session$ and happened in the last 1 hour. Formally, this set of history sessions denoted by $\Agg{\Model^{\optimal}_\session}{\session}$.
%\item Finally, \name predicts the throughput of $\session$ by $\Pred{\session}=\QualitySummary(\Agg{\Model^{\optimal}_\session}{\session})$, where $\QualitySummary(S)$ reports the median of the throughput of sessions in $S$.
%\end{packedenumerate}
%
%The goal of $\Pred{\session}$ is to minimize the prediction error $\Error{\Pred{\session}}{\Throughput{\session}}=\frac{|\Pred{\session}-\Throughput{\session}|}{\Throughput{\session}}$, where  $\Throughput{\session}$ is the actual throughput of $\session$

{\footnotesize
\vspace{-0.4cm}
\begin{align}
&\Model^{\optimal}_\session=\argmin_{\Model}
\Error{\QualitySummary(\Agg{\Model}{\session})}{\Throughput{\session}} \label{eq:goal}
\end{align}
\vspace{-0.4cm}
}

\noindent Rather than solving Eq~\ref{eq:goal} analytically, \name takes a {\em data-driven} approach and finds the best prediction model over a set of history sessions $\Estimation{\session}$ (defined shortly). Formally, the process can be written as following:

{\footnotesize
\vspace{-0.4cm}
\begin{align}
&\Model^{\optimal}_\session=\argmin_{\Model}\frac{1}{|\Estimation{\session}|}\sum_{\session'\in\Estimation{\session}}
\Error{\QualitySummary(\Agg{\Model}{\session'})}{\Throughput{\session'}} \label{eq:empirical}
\end{align}
\vspace{-0.4cm}
}

\noindent$\Estimation{\session}$ should include sessions that are likely to share the best prediction model with $\session$. In \name, $\Estimation{\session}$ consists of sessions that match features $\mathit{Target}$, $\mathit{ISP}$, $\mathit{Technology}$ and $\mathit{Downlink}$ with $\session$ and happened within 4 hours before $\session$.

\myparatight{Estimating throughput} Second, \name estimates $\session$'s throughput by the learned prediction model $\Model^{\optimal}_\session$.
%\name also adopts two fine-grain optimizations to further improve corner cases. First, 
To make the prediction $\Pred{\session}$ reliable, \name ensures that $\Pred{\session}$ is based on a substential amount of sessions in $\Agg{\Model^{\optimal}_\session}{\session}$. Therefore, if $\Model^{\optimal}_\session$ yields $\Agg{\Model^{\optimal}_\session}{\session}$ with less than 20 sessions, \name will remove that model from the pool and learn the prediction model as in the first step again.
%Second, 
We have also found that for some pairs of client and server, \name's prediction error is one-sided. For instance, the throughput of a particular client-server pair is 1Mbps, while the best prediction model always predicts 2Mbps (i.e., a one-sided 100\% error). We compensate this error by changing $\QualitySummary(S)$ to $\QualitySummary(S,k)$ which reports the median of throughput in $S$ times a factor $k$. To train a proper value of $k$, \name first uses Eq~\ref{eq:empirical} to learn $\Model^{\optimal}_\session$ by assuming $k=1$, and then, \name trains the best factor $k^{\optimal}_\session$ for $\session$ as follows: 

{\footnotesize
\vspace{-0.4cm}
\begin{align}
&k^{\optimal}_\session=\argmin_{k}\frac{1}{|\Estimation{\session}|}\sum_{\session'\in\Estimation{\session}}
\Error{\QualitySummary(\Agg{\Model^{\optimal}_\session}{\session'},k)}{\Throughput{\session'}} \nonumber
\end{align}
\vspace{-0.4cm}
}

\noindent where $k$ is chosen from 0 to 5. Finally, the prediction made by \name will be $\QualitySummary(\Agg{\Model^{\optimal}_\session}{\session},k^{\optimal}_\session)$.

\tightsection{Evaluation}
\label{sec:eval}

This section evaluates the prediction accuracy of \name (\Section\ref{subsec:accuracy-eval}) and how much \name improves video bitrate (\Section\ref{subsec:bitrate-eval}). Overall, our findings show the following: 
\begin{packedenumerate}
\item \name can predict more accurately than other predictors.
\item With higher accuracy, \name can select better bitrate.
\end{packedenumerate}

\tightsubsection{Prediction accuracy}
\label{subsec:accuracy-eval}

\myparatight{Methodology} As points of comparison, we use implementations of Decision Tree (DT) and Naive Bayes (NB) with default configurations in \texttt{weka}, a popular ML tool~\cite{weka}. For a fair comparison, all algorithms use the same set of features. We also compare them with last-mile predictor (LM)\footnote{LM is not applicable to the VoD dataset as it has no feature related to last-mile connection.} and last-sample predictor (LS), introduced in \Section\ref{sec:analysis}. We update the model of other algorithms in a same way as \name: for each session under prediction, we use all available history data before it as the train data. Each session's timestamp is grouped into 10-minute intervals and used as discrete time feature. By default, we use absolute normalized error (\Section\ref{sec:algorithm}) as the metric of prediction error, and the results are based on the FCC dataset, unless specified otherwise.

\begin{figure}[h!]
\centering
\hspace{-0.6cm}
\subfigure[FCC]
{
        \includegraphics[width=0.23\textwidth]{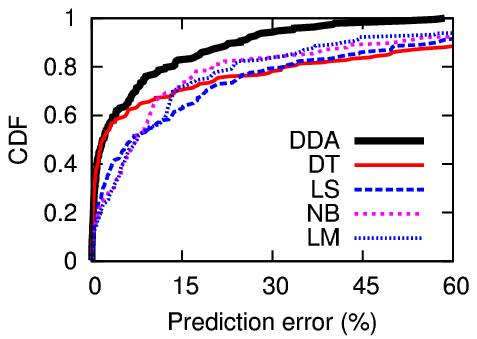}
        \label{fig:prediction-error-fcc}
}
\hspace{-0.6cm}
\subfigure[VoD China]
{
        \includegraphics[width=0.23\textwidth]{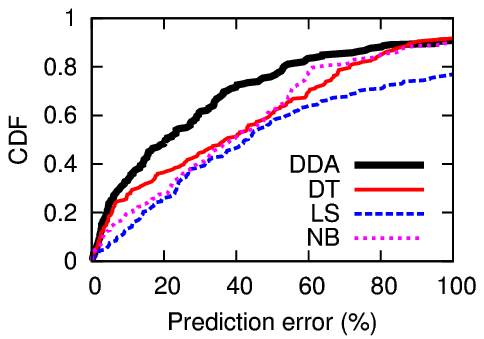}
        \label{fig:prediction-error-pptv}
}
\hspace{-0.6cm}
\tightcaption{CDF of prediction error.}
\label{fig:prediction-error}
\end{figure}

\myparatight{Distribution of prediction error} Figure~\ref{fig:prediction-error} shows the distribution of prediction error of \name and other algorithms.
\name outperforms all algorithms, especially on the tail of prediction error. For the FCC dataset (Figure~\ref{fig:prediction-error-fcc}), 80\%ile prediction error of \name is 50\% to 80\% lower than that of other algorithms, and \name has less than 20\% sessions with more than 10\% prediction error, while all other algorithms have at least 30\% session with more than 10\% error. While the VoD dataset in general has higher prediction error than the FCC dataset (due to the lack of some features such as last-connection and longitudinal information), \name still outperforms other algorithms, showing that \name is robust to the available features.

\begin{figure*}[t!]
\centering
\hspace{-0.6cm}
\subfigure[Prediction error vs. ISP]
{
        \includegraphics[width=0.3\textwidth]{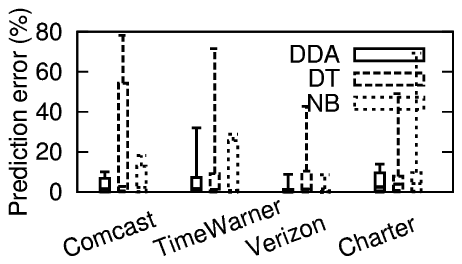}
        \label{fig:insight-isp}
}
%\hspace{-0.6cm}
\subfigure[Prediction error vs. time of day (UTC)]
{
        \includegraphics[width=0.3\textwidth]{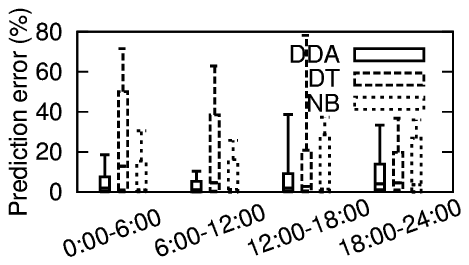}
        \label{fig:insight-timeofday}
}
%\hspace{-0.6cm}
\subfigure[Prediction error vs. random drop rate]
{
        \includegraphics[width=0.3\textwidth]{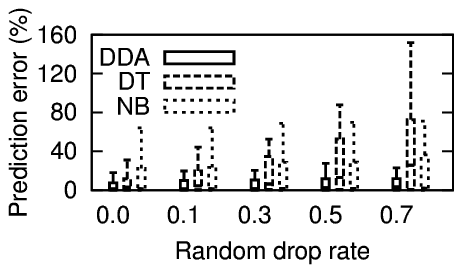}
        \label{fig:sensitivity}
}
\hspace{-0.6cm}
\tightcaption{Dissecting \name prediction error. The boxes show the 10-20-50-80-90 percentile.}
\label{fig:insight}
\end{figure*}

\myparatight{Dissecting prediction accuracy of \name} To evaluate the prediction accuracy in more details, we first partition the prediction error by four most popular ISPs (\ref{fig:insight-isp}) and by different time of day (\ref{fig:insight-timeofday}). Although the ranking of algorithms varies across different partitions, \name consistently outperforms other two algorithms (DT, NB), especially in the tail of 90\%ile. Finally, Figure~\ref{fig:sensitivity} evaluates \name's sensitivity to measurement frequency by comparing the distribution of prediction error of three algorithms under different random drop rates (\Section\ref{sec:dataset}). It shows that \name is more robust to measurement frequency than the other algorithms.
%When random drop rate is zero, the measurement frequency between each client and target is once every hour, and when random drop rate is 0.5, the measurement frequency between each client and target is in average once every 2 hours. 

%the FCC dataset has one measurement between a client and target every hour, and this allows us to evaluate the prediction accuracy under different measurement frequency. To do this, we randomly drop part of the measurement in the FCC dataset, and Figure~\ref{fig:sensitivity} shows the average prediction error and their distribution of three prediction algorithms under different random drop rate. When random drop rate is zero, the measurement frequency between each client and target is once every hour, and when random drop rate is 0.99, the measurement frequency between each client and target is in average once every 100 hours. First, Figure~\ref{fig:sensitivity-mean} shows that the average prediction error of \name is more robust to measurement frequency than the other algorithms (NN and NB). Second, Figure~\ref{fig:sensitivity-distribution} shows that the tails (10,20,80,90 percentile) of prediction error \name is much more stable than the other algorithms against measurement frequency.

\tightsubsection{Improvement of bitrate selection}
\label{subsec:bitrate-eval}

%This section shows the bitrate improvement resulted from accuracy throughput prediction of \name. 

\myparatight{Methodology} To evaluate bitrate selected based on some prediction algorithm, we consider a simple bitrate selection algorithm (while a more complex algorithm is possible, it is not the focus of this paper): given a session of which the prediction algorithm predicts the throughput by $w$, the bitrate selection algorithm simply picks highest bitrate from \{0.016, 0.4, 1.0, 2.5, 5.0, 8.0, 16.0, 35.0\}Mbps~\cite{youtube-bitrates,wiki-bitrates} and below $\alpha w$, where $\alpha$ represents the safety margin (e.g.,  higher $\alpha$ means higher bitrate at the risk of exceeding the throughput). We use two metrics to evaluate the performance: (1) AvgBitrate -- average value of picked bitrate, and (2) GoodRatio -- percentage of sessions with no re-buffering (i.e., picked bitrate is lower than the throughput). Therefore, one bitrate selection algorithm is better than another if it has both higher AvgBitrate and higher GoodRatio.
As points of reference, ``Global'' bitrate selection algorithm picks the same bitrate for any session, which represents how today's players select starting bitrate. As a optimal reference point, ``Ideal'' bitrate selection algorithm picks the bitrate identical to the throughput for any session (\Section\ref{sec:dataset}).

\myparatight{Overall improvement} Table~\ref{tab:bitrate-overall} compares \name-based bitrate selection and the ``Global''. In both algorithms, we use $\alpha=0.8$ for the FCC dataset, and $\alpha=0.6$ for the VoD dataset. In both datasets, \name leads to higher AvgBitrate and GoodRatio, and \name is much closer to ``Ideal'' than ``Global''. Note that the VoD dataset still has a substantial room of improvement due to the relatively low prediction accuracy (Figure~\ref{fig:prediction-error-pptv}).

\begin{table}[h]
\begin{footnotesize}
\begin{tabular}{l|l|l|l|l}
       & \multicolumn{2}{c}{{\bf FCC}} & \multicolumn{2}{c}{{\bf VoD China}} \\ 
       & {\bf AvgBitrate} & {\bf GoodRatio} & {\bf AvgBitrate} & {\bf GoodRatio} \\\hline\hline
Global &    2.5Mbps       &  88.2\%         & 2.5Mbps          &  77.5\%           \\
\name  &    13.3Mbps      &  99.5\%         & 2.7Mbps          &  88.2\%           \\
Ideal  &    27.2Mbps      &  100\%          & 3.5Mbps          &  100\%              
\end{tabular}
\end{footnotesize}
\tightcaption{Comparing \name and ``Global'' in AvgBitrate and GoodRatio.}
\label{tab:bitrate-overall}
\end{table}

\myparatight{Bitrate selection vs. prediction accuracy} Next, we examine the intuition that higher prediction accuracy leads to higher performance of bitrate selection. Table~\ref{tab:bitrate-accuracy} shows the bitrate selection performance as a function of median prediction error. We consider four prediction algorithms (\name, DT, LS, NB). For a fair comparison, the bitrate selection algorithm always uses $\alpha=0.8$. As prediction error increases, the performance of bitrate selection degrades in terms of both lower AvgBitrate and lower GoodRatio.

%\begin{figure}[h!]
%\centering
%\hspace{-0.6cm}
%\subfigure[Accuracy vs. AvgBitrate]
%{
%        \includegraphics[width=0.2\textwidth]{figures/bitrate-accuracy-bitrate.eps}
%        \label{fig:bitrate-accuracy-bitrate}
%}
%%\hspace{-0.6cm}
%\subfigure[Accuracy vs. GoodRatio]
%{
%        \includegraphics[width=0.2\textwidth]{figures/bitrate-accuracy-good.eps}
%        \label{fig:bitrate-accuracy-good}
%}
%\hspace{-0.6cm}
%\tightcaption{Higher accuracy means better bitrate selection.}
%\label{fig:bitrate-accuracy}
%\end{figure}

\begin{table}[h]
\begin{footnotesize}
\begin{tabular}{p{0.8cm}|p{2.3cm}|p{1.5cm}|p{1.5cm}}
    & {\bf Mean/median prediction error} & {\bf AvgBitrate} & {\bf GoodRatio} \\ \hline\hline
\name & 9.0\%/2.3\%    & 13.3Mbps              & 99.5\%           \\
DT  & 23.1\%/3.4\%    & 13.0Mbps              & 91.0\%           \\
LS  & 28.7\%/9.8\%    & 12.3Mbps              & 90.6\%           \\
NB  & 91.4\%/17.1\%    & 12.2Mbps              & 71.8\%          
\end{tabular}
\end{footnotesize}
\tightcaption{Higher accuracy means better bitrate selection.}
\label{tab:bitrate-accuracy}
\end{table}

\begin{figure}[h!]
\centering
\hspace{-0.6cm}
\subfigure[AvgBitrate-GoodRatio tradeoff]
{
        \includegraphics[width=0.24\textwidth]{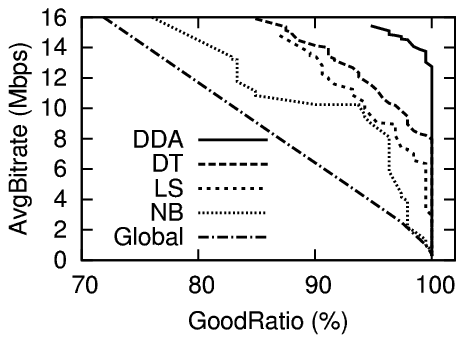}
        \label{fig:bitrate-tradeoff}
}
%\hspace{-0.6cm}
\subfigure[Performance by ISP]
{
        \includegraphics[width=0.23\textwidth]{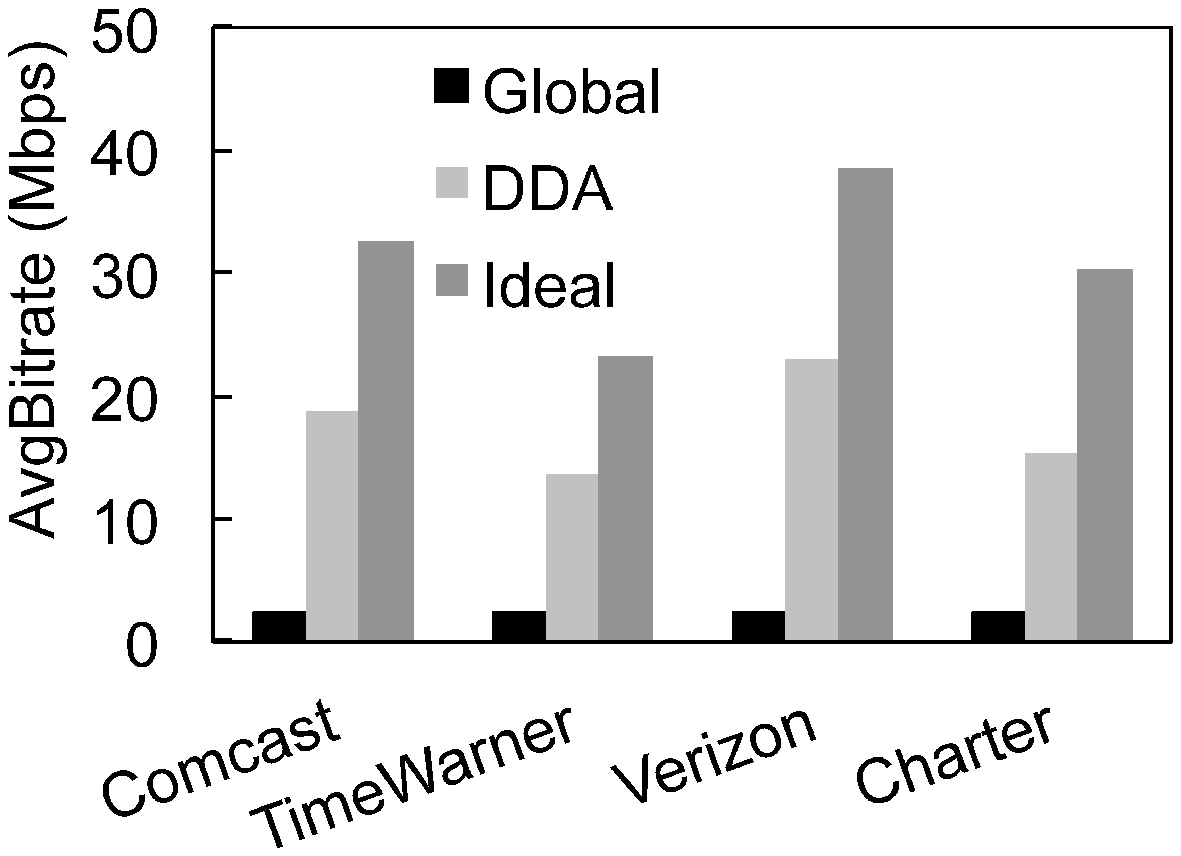}
        \label{fig:bitrate-partition-isp}
}
\hspace{-0.6cm}
\tightcaption{In-depth analysis of bitrate selection}
\label{fig:bitrate-indepth}
\end{figure}

\myparatight{Understanding bitrate improvement} There is a natural tradeoff between AvgBitrate and GoodRatio (e.g., higher $\alpha$ means higher AvgBitrate at the cost of lower GoodRatio). Figure~\ref{fig:bitrate-tradeoff} shows such tradeoff of various bitrate selection algorithms by adjusting the value $\alpha$. It is shown that \name-based bitrate selection strikes a better tradeoff of higher AvgBitrate and higher GoodRatio (i.e., more towards the top-right corner of the figure). 

Finally, we would like to test the robustness of \name-based bitrate selection in different regions. Figure~\ref{fig:bitrate-partition-isp} compares the AvgBitrate of \name with ``Global'' and ``Ideal'' in four popular ISPs. \name uses the maximum $\alpha$ on the tradeoff curve in Figure~\ref{fig:bitrate-tradeoff} that ensures at least 95\% GoodRatio, while ``Global'' only has GoodRatio of 88.2\%. 
Across all ISPs, \name consistently outperforms ``Global'' and achieve at least 60\% of the ``Ideal''.
%Comcast and Verizon have a large room of improvement (i.e., large gap between ``Global'' and ``Ideal''), and \name-based bitrate selection achieves close-to-ideal performance, resulting a higher improvement on these two ISPs. In contrast, there is substantial gap between \name and ``Ideal'' on Time Warner Cable and Charter, due to relatively higher prediction error (see Figure~\ref{fig:insight-isp}).

%\section{Limitation of this work}
%
%\begin{itemize}
%\item we don't consider scalability, though it could be addressed potentially using the same insights in section 2.
%\item we don't consider intra-session throughput variability.
%\end{itemize}

\section{Conclusion}
Many Internet applications can benefit from estimating end-to-end throughput.  This paper focuses on its application to initial video bitrate selection. We present \name, which leverages the throughput measured by different clients and servers to achieve accurate throughput prediction before a new session starts. Evaluation based on two real-world datasets shows (i) \name predicts throughput more accurately than simple predictors and conventional machine learning algorithms, and (ii) with more accurate throughput prediction, a player can choose a higher-yet-sustainable bitrate (e.g., compared to initial bitrate without prediction, \name leads to $4\times$ higher average bitrate with less sessions using bitrate exceeding the throughput).

\newcommand\blfootnote[1]{%
  \begingroup
  \renewcommand\thefootnote{}\footnote{#1}%
  \addtocounter{footnote}{-1}%
  \endgroup
}
%\newpage
%{
%\footnotesize
%\bibliographystyle{abbrv}
%\bibliography{}
%}
{\scriptsize
\bibliographystyle{abbrv}
\bibliography{adaptation,sigcomm2011,sigcomm2012,sigcomm2013,conext13,sigcomm2014,nsdi13,hotnet14,nsdi14,sigcomm2015,imc2015.bib}

\begin{thebibliography}{10}

\bibitem{fcc-methodology}
2014 measuring broadband america report technical appendix.
\newblock
  {\url{http://data.fcc.gov/download/measuring-broadband-america/2014/Technical-Appendix-fixed-2014.pdf}}.

\bibitem{wiki-bitrates}
Bitrates in multimedia.
\newblock {\url{http://en.wikipedia.org/wiki/Bit_rate#Bitrates_in_multimedia}}.

\bibitem{informationgain}
Information gain.
\newblock {\url{http://www.autonlab.org/tutorials/infogain11.pdf}}.

\bibitem{fcc-2014}
Measuring broadband america 2014.
\newblock
  {\url{https://www.fcc.gov/measuring-broadband-america/2014/validated-data-fixed-2014}}.

\bibitem{netflix}
Netflix.
\newblock \url{www.netflix.com/}.

\bibitem{youtube-bitrates}
Recommended upload encoding settings.
\newblock {\url{https://support.google.com/youtube/answer/1722171?hl=en}}.

\bibitem{weka}
The weka manual 3.6.10.
\newblock {\url{http://goo.gl/ISSY3c}}.

\bibitem{dash}
{ I. Sodagar}.
\newblock { The MPEG-DASH Standard for Multimedia Streaming Over the Internet}.
\newblock {\em IEEE Multimedia}, 2011.

\bibitem{sigcomm13athula}
A.~Balachandran, V.~Sekar, A.~Akella, S.~Seshan, I.~Stoica, and H.~Zhang.
\newblock Developing a predictive model of quality of experience for internet
  video.
\newblock In {\em ACM SIGCOMM '13}.

\bibitem{balakrishnan1997analyzing}
H.~Balakrishnan, M.~Stemm, S.~Seshan, and R.~H. Katz.
\newblock Analyzing stability in wide-area network performance.
\newblock In {\em ACM SIGMETRICS Performance Evaluation Review}. ACM, 1997.

\bibitem{dabek2004vivaldi}
F.~Dabek, R.~Cox, F.~Kaashoek, and R.~Morris.
\newblock Vivaldi: A decentralized network coordinate system.
\newblock In {\em In SIGCOMM}, 2004.

\bibitem{he2005predictability}
Q.~He, C.~Dovrolis, and M.~Ammar.
\newblock On the predictability of large transfer tcp throughput.
\newblock {\em ACM SIGCOMM Computer Communication Review}, 35(4):145--156,
  2005.

\bibitem{hu2005measurement}
N.~Hu, L.~Li, Z.~M. Mao, P.~Steenkiste, and J.~Wang.
\newblock A measurement study of internet bottlenecks.
\newblock In {\em INFOCOM 2005. 24th Annual Joint Conference of the IEEE
  Computer and Communications Societies. Proceedings IEEE}, volume~3, pages
  1689--1700. IEEE, 2005.

\bibitem{hu2004locating}
N.~Hu, L.~E. Li, Z.~M. Mao, P.~Steenkiste, and J.~Wang.
\newblock Locating internet bottlenecks: Algorithms, measurements, and
  implications.
\newblock In {\em In Proc. of ACM SIGCOMM’04}, 2004.

\bibitem{huang2014buffer}
T.-Y. Huang, R.~Johari, N.~McKeown, M.~Trunnell, and M.~Watson.
\newblock A buffer-based approach to rate adaptation: evidence from a large
  video streaming service.
\newblock In {\em ACM SIGCOMM 2014}.

\bibitem{jain2003end}
M.~Jain and C.~Dovrolis.
\newblock End-to-end available bandwidth: measurement methodology, dynamics,
  and relation with tcp throughput.
\newblock {\em IEEE/ACM Transactions on Networking (TON)}, 11(4):537--549,
  2003.

\bibitem{jain2005end}
M.~Jain and C.~Dovrolis.
\newblock End-to-end estimation of the available bandwidth variation range.
\newblock In {\em ACM SIGMETRICS Performance Evaluation Review}, volume~33,
  pages 265--276. ACM, 2005.

\bibitem{festive}
J.~Jiang, V.~Sekar, and H.~Zhang.
\newblock {Improving Fairness, Efficiency, and Stability in HTTP-Based Adaptive
  Streaming with Festive }.
\newblock In {\em ACM CoNEXT 2012}.

\bibitem{madhyastha2006iplane}
H.~Madhyastha, T.~Isdal, M.~Piatek, C.~Dixon, T.~Anderson, A.~Krishnamurthy,
  and A.~Venkataramani.
\newblock iplane: An information plane for distributed services.
\newblock In {\em Proceedings of the 7th symposium on Operating systems design
  and implementation}, pages 367--380. USENIX Association, 2006.

\bibitem{miller2015low}
K.~Miller, A.-K. Al-Tamimi, and A.~Wolisz.
\newblock Low-delay adaptive video streaming based on short-term tcp throughput
  prediction.
\newblock 2015.

\bibitem{mirza2007machine}
M.~Mirza, J.~Sommers, P.~Barford, and X.~Zhu.
\newblock A machine learning approach to tcp throughput prediction.
\newblock In {\em ACM SIGMETRICS Performance Evaluation Review}, volume~35,
  pages 97--108. ACM, 2007.

\bibitem{nikravesh2014mobile}
A.~Nikravesh, D.~R. Choffnes, E.~Katz-Bassett, Z.~M. Mao, and M.~Welsh.
\newblock Mobile network performance from user devices: A longitudinal,
  multidimensional analysis.
\newblock In {\em Passive and Active Measurement}, pages 12--22. Springer,
  2014.

\bibitem{prasad2003bandwidth}
R.~Prasad, C.~Dovrolis, M.~Murray, and K.~Claffy.
\newblock Bandwidth estimation: metrics, measurement techniques, and tools.
\newblock In {\em Network, IEEE}. IEEE, 2003.

\bibitem{ramasubramanian2009treeness}
V.~Ramasubramanian, D.~Malkhi, F.~Kuhn, M.~Balakrishnan, A.~Gupta, and
  A.~Akella.
\newblock On the treeness of internet latency and bandwidth.
\newblock In {\em ACM SIGMETRICS Performance Evaluation Review}, volume~37,
  pages 61--72. ACM, 2009.

\bibitem{seshan1997spand}
S.~Seshan, M.~Stemm, and R.~H. Katz.
\newblock Spand: Shared passive network performance discovery.
\newblock In {\em USENIX Symposium on Internet Technologies and Systems}, pages
  135--146, 1997.

\bibitem{strauss2003measurement}
J.~Strauss, D.~Katabi, and F.~Kaashoek.
\newblock A measurement study of available bandwidth estimation tools.
\newblock In {\em Proceedings of the 3rd ACM SIGCOMM conference on Internet
  measurement}, pages 39--44. ACM, 2003.

\bibitem{swany2002multivariate}
M.~Swany and R.~Wolski.
\newblock Multivariate resource performance forecasting in the network weather
  service.
\newblock In {\em Proceedings of the 2002 ACM/IEEE conference on
  Supercomputing}, pages 1--10. IEEE Computer Society Press, 2002.

\bibitem{tian2012towards}
G.~Tian and Y.~Liu.
\newblock Towards agile and smooth video adaptation in dynamic http streaming.
\newblock In {\em Proceedings of the 8th international conference on Emerging
  networking experiments and technologies}, pages 109--120. ACM, 2012.

\bibitem{vazhkudai2001predicting}
S.~Vazhkudai, J.~M. Schopf, and I.~Foster.
\newblock Predicting the performance of wide area data transfers.
\newblock In {\em Parallel and Distributed Processing Symposium., Proceedings
  International, IPDPS 2002, Abstracts and CD-ROM}, pages 10--pp. IEEE, 2001.

\bibitem{yin2014toward}
X.~Yin, V.~Sekar, and B.~Sinopoli.
\newblock Toward a principled framework to design dynamic adaptive streaming
  algorithms over http.
\newblock In {\em ACM HotNets}, 2014.

\bibitem{zhang2001constancy}
Y.~Zhang and N.~Duffield.
\newblock On the constancy of internet path properties.
\newblock In {\em Proceedings of the 1st ACM SIGCOMM Workshop on Internet
  Measurement}, pages 197--211. ACM, 2001.

\end{thebibliography}
}

%\newpage
%\appendix
%\input{todo}

\end{document}